# Simulation of the Spin Field Effect Transistors:

# Effects of Tunneling and Spin Relaxation on its Performance

Yunfei Gao[1], Tony Low[1], Mark S. Lundstrom[1], and Dmitri E. Nikonov[2]

[1]Department of Electrical and Computer Engineering, Purdue University, West Lafayette, Indiana 47906, USA

[2]Components Research, Intel Corporation, 2200 Mission College Blvd., Santa Clara, California 95052, USA

## Abstract

A numerical simulation of spin-dependent quantum transport for a spin field effect transistor (spinFET) is implemented in a widely used simulator nanoMOS. This method includes the effect of both spin relaxation in the channel and the tunneling barrier between the source/drain and the channel. Account for these factors permits setting more realistic performance limits for the transistor, especially the magnetoresistance, which is found to be lower compared to earlier predictions. The interplay between tunneling and spin relaxation is elucidated by numerical simulation. Insertion of the tunneling barrier leads to an increased magnetoresistance. Numerical simulations are used to explore the tunneling barrier design issues.

# 1. Introduction

Spin-based logic is currently being explored as one of the beyond-CMOS computing technologies [1], which are presently being considered to supplement complementary metal-oxide-semiconductor (CMOS) field effect transistors (FET) in microprocessors. Vigorous research in spintronic devices has been carried out over the last two decades [2, 3, 4] and has resulted in demonstration of two-terminal giant magnetoresistance (GMR) [5, 6] and tunneling magnetoresistance (TMR) [7] devices, and their switching by spin transfer torque [8]. Such devices have been commercialized in magnetic hard drives and magnetic RAMs, and had a great impact on every-day life. The question arises now whether there is a spintronic device capable of similar success in logic.

One of the candidates is the spin field effect transistor (spinFET) first proposed by Sugahara and Tanaka [9], a three-terminal device that utilizes ferromagnetic contacts in the source and drain as spin injector and detector. The spinFET is essentially a combination of two Schottky barrier MOSFETs, each implemented by carriers with a certain spin state (e.g. one up-spin and one down-spin). The transport channels for up-spin and down-spin electrons (or holes) are independent, if no spin relaxation occurs, but they become interconnected if spin flip processes happen. The semiconductor channel makes the spinFET compatible with the modern CMOS technology. Relatively small spin orbit coupling and negligible hyperfine interaction gives electrons a long spin life time in silicon [10], which makes it a good candidate for the channel material. However, spinFETs are also envisioned with germanium or III-V semiconductor channels. Due to the exchange splitting between the up-spin and down-spin bands in the ferromagnetic contacts, the up-spin and down-spin electrons experience different Schottky barriers from their conduction bands as they enter into and escape out of the channel. The gate controls the width of these Schottky barriers and the electrostatic potential in the channel. The magnetizations of the source and drain can be switched to be parallel or anti-parallel to obtain low or high and low resistance between these contacts, respectively, similarly to a magnetic tunnel junctions (MTJ) [7, 11]. Therefore, the current flow is controlled by the gate and drain bias, and also by the direction of the contacts' magnetization. The switching of magnetization can be performed, for example, by spin transfer torque of the flowing current. SpinFET must be distinguished from the spin modulator based on spin precession, the original spintronic device proposed by Datta and Das [12]. We will not consider this device here, though some later publications call it "spin field effect transistor".

The magnetoresistance (MR) ratio, which is a key device performance metric of a spintronic device, is defined via the resistances in parallel ($R_P$) and anti-parallel ($R_{AP}$) contact magnetization configurations as follows $MR=(R_{AP}-R_P)/R_P$. The identical quantity (provided a fixed voltage is applied) is magnetocurrent ratio $MC=(I_P-I_{AP})/I_{AP}$. In order to improve MR, high spin polarization in both source and drain contacts is favorable. Half-metal ferromagnets (HMF) were predicted [13], and later on demonstrated by experiment [14, 15], to have close to 100% spin polarization of electrons, which is desirable for the contact ferromagnetic material. With the ideal performance of spinFETs, it is further shown in [16] that non-volatile memory and reconfigurable logic circuits can be constructed using these devices. Despite of the theoretically predicted perfect spin polarization in the bulk HMF, there has been no observation of high spin-polarized current injected from the HMF in experiments [17]. It is argued that when the HMF gets in contact with the non-magnets, a randomization layer is formed at the interface [18] where spins of electrons are not aligned. This inevitable non-ferromagnetic layer can decrease the injected polarization and reduce the MR ratio [19].

Conduction mismatch between the ferromagnet and the semiconductor is another reason for the non-ideal spin injection from [20]. The solution was found in inserting a tunneling barrier between the ferromagnet and the semiconductor [21,22]. Even though the tunneling barrier resistance decreases the current, a significant enhancement of the injection efficiency is obtained. The third factor for non-ideality of spinFET is the spin flip (SF) scattering in the channel. In the presence of SF scattering, the two conducting channels (up-spin and down-spin) are mixed, which has a great impact on the carriers transport and the MR ratio. All these unavoidable imperfection of spinFETs should be taken into account when simulating the devices and assessing their performance potential.

An experimental pre-requisite to spinFET is not just electrical injection of spin polarization in a semiconductor, but also electrical detection of spin polarization [23,24]. Necessary conditions for efficient spin injection-detection and high MR have been determined theoretically [22,25]. One of them is low-resistance tunneling interface between the FM and a semiconductor. Low-resistance interface to Si [26] and Ge [27] have been fabricated and characterized. A spinFET has been demonstrated only recently [28]; it contained HMF electrodes and was switched by spin torque effect.

There has been a large number of theoretical and simulation studies on spin injection from ferromagnets into semiconductors, see review [29]. Spin injection to semiconductors has been studied in a classical approximation, with drift-diffusion type of equations [30]. The Non-equilibrium Green's function (NEGF) method [31] is a rigorous quantum transport treatment of nano-scale devices. First the NEGF method has been applied in the research of MTJ devices [32, 33]. NEGF based on the density functional theory has also been used to study MTJs [34, 35]. A spinFET was treated by NEGF [19], however the transport in the channel was considered as ballistic with relaxation only at the source/drain and channel interfaces.

The present article reports the following advances compared to the prior work: (1) simulation of spin-dependent quantum transport in a ferromagnet-semiconductor ferromagnet structure, including tunneling barriers, (2) rigorous treatment of spin scattering, both in the channel and the randomization layer, (3) set realistic performance limits (especially MR) for spinFET with relevant factors of non-ideality, (4) implementation within a well established quantum transport simulator, nanoMOS [36].

The rest of this paper is structured as follow. In section II, we summarize the NEGF formalism used to describe the carrier transport in spinFETs, and more specifically focus on the mathematical treatment of SF scatterings and the physical connection with spin lifetime in various materials. In section III, we apply this method to realistically structured spinFETs and quantitatively show that the SF scattering affects the I-V characteristics and can dramatically reduce the MR ratio. The physical reasoning is then presented along with rigorous simulation results, and two solutions to enhance the MR ratio are proposed and examined by numerical simulation. Finally conclusions are drawn in section IV.

## 2. Numerical model

The NEGF formalism is ideally suited for analyzing quantum transport of carriers in nanoscale devices. In this section we first briefly restate the main equations of the NEGF method necessary for understanding the results. For a more detailed presentation see [31,37]. Then we apply it to the case of a spinFET with spin scattering. The key numerical model is described and connection of the mathematical description with the physical model is discussed.

## 2.1 NEGF method

The channel material is described by a Hamiltonian matrix $[H]$ of the size $N \times N$ blocks, $N$ being the total number of grid points in the transport direction. Charging effect due to the interaction of carrier charges with the potential of the rest of the charges is incorporated via the potential matrix $[U]$. These serve as inputs in the equation for the retarded Green's function at a specific value of energy $E$:

$$G(E) = \left[EI - H - U - \Sigma(E)\right]^{-1} \qquad (1)$$

The self-energy accounts for non-coherent processes and contains terms due to the left and right contacts and due to scattering processes in the device

$$\Sigma(E) = \Sigma_L(E) + \Sigma_R(E) + \Sigma_S(E) \qquad (2)$$

And similarly, the in- and out-scattering functions describe the change in populations of electrons and holes due to these incoherent processes

$$\Sigma^{in/out}(E) = \Sigma_L^{in/out}(E) + \Sigma_R^{in/out}(E) + \Sigma_S^{in/out}(E) \tag{3}$$

The spectral function [A] related to the local density of states, and the electron/hole correlation function $[G^{n/p}]$, proportional to the occupation numbers of electrons and holes in states of certain energyare given by

$$A(E) = i[G(E) - G^\dagger(E)] \tag{4}$$

$$G^{n/p}(E) = G(E)\Sigma^{in/out}(E)G^\dagger(E) \tag{5}$$

They are related to the local density of states, so they also satisfy the relation

$$A(E) = G^n(E) + G^p(E) \tag{6}$$

The strength of coupling to the left (source) and right (drain) contacts are described by the broadening matrices which are related to imaginary parts of the corresponding self-energy matrices:

$$\Gamma_{L/R}(E) = i[\Sigma_{L/R}(E) - \Sigma_{L/R}^\dagger(E)] \tag{7}$$

The in-scattering/out-scattering matrices represent the carrier injection and extraction rates into/out of the channel:

$$\Sigma_{L/R}^{in}(E) = f_{L/R}(E)\Gamma_{L/R}(E), \quad \Sigma_{L/R}^{out}(E) = [1 - f_{L/R}(E)]\Gamma_{L/R}(E) \tag{8}$$

where $f_{L/R}(E)$ is the Fermi distribution functions in each contact.

Scattering, no matter if it is elastic or inelastic, can be visualized as the coupling of the channel and a reservoir [31], whose self-energy matrix is $[\Sigma_S]$. The scattering process is physically described by the in-scattering/out-scattering matrices $[\Sigma_S^{in/out}]$ which show the rate of electrons coming into/out of a certain state. The sum of the two matrices gives the broadening matrix due to the scattering process:

$$\Gamma_S(E) = \Sigma_S^{in}(E) + \Sigma_S^{out}(E) \qquad (9)$$

from which the scattering-related self-energy can be obtained through a Hilbert transform as

$$\Sigma_S(E) = P\left[\int \frac{dE'\Gamma_S(E')}{2\pi(E-E')}\right] - i\frac{\Gamma_S(E)}{2} \qquad (10)$$

The imaginary part of $[\Sigma_S]$ obeys the same fashion as that between $[\Gamma_{L/R}]$ and $[\Sigma_{L/R}]$. The real part of $[\Sigma_S]$ is obtained via the Hilbert transform, where *P* stands for the principal value of a singular integral, see [38] for details.

The NEGF and the Poisson equation are solved self-consistently in a loop, because the electron density is obtained from NEGF equations and used to solve for the electric potential, while the potential is necessary to solve the NEGF equations. The current is calculated once the consistency is reached. This is the only loop necessary for the ballistic simulation (i.e. with zero scattering terms). In the scattering case, we have to consider an additional inner self-consistency loop to calculate the in-scattering or out-scattering matrix $[\Sigma_S^{in/out}]$ and the electron/hole correlation function $[G^{n/p}]$ in the NEGF formalism. As described in [31,37], the in-/out-scattering energies $[\Sigma_S^{in/out}]$ contains $[G^{n/p}]$ as the inputs. They are used, in their turn, to calculated the contact self-energy [$\Sigma_S$] through Eq. (9) and (10), and consequently to obtain [$G$] from Eq. (1). Once the self-consistency in the inner loop is achieved, the iteration in the outer loop of NEGF and Poisson equations starts. One way to speed up the simulation is to bypass the computationally intensive Hilbert transform in Eq. (9). It is possible for elastic scattering, where the in-/out-scattering functions depend on the Green's functions at the same value of energy only. In that case, the expression of the self-energy drastically simplifies, see [38]. The spin flip (SF) scattering considered here is elastic, and thus admits such a simplification. Thus the expressions for the scattering terms become

$$\Sigma_S(E) = \mathbf{D}(E)G(E), \quad \Sigma^{in}(E) = \mathbf{D}(E)G^n(E), \quad \Sigma^{out}(E) = \mathbf{D}(E)G^p(E), \quad \Gamma(E) = \mathbf{D}(E)A(E)$$

(11)

where we introduced the scattering tensor [$\mathbf{D}$]. In this case, a simpler self-consistency loop is performed to calculate the Green's functions at separate values of energy, which proves to be less time consuming.

At node $i$ of the grid, total current ($I_i$) and current for each energy level ($I_i(E)$) are given by the summation over spin states $I$ and the integral over energies.

$$\tilde{I}_i(E) = \sum_s \frac{ie}{\hbar} \left[ H_{i,i+1} G^n_{i+1,i}(E) - H_{i+1,i} G^n_{i,i+1}(E) \right] \quad (12)$$

$$I_i = \int_{-\infty}^{+\infty} \frac{dE}{2\pi} \tilde{I}_i(E) \quad (13)$$

## 2.2 SpinFET description

The structure of the spinFET is illustrated in Fig. 1. Current flows along the transport direction *x*. Across the *z* direction are two metal gates separated by thin dielectric layers of gate oxide above and below the channel which provide good electrostatics control. We have implemented the spin-dependent transport simulation based on the widely used simulator nanoMOS [36]. The width of the device in the transverse direction *z* is assumed to be large enough, so that the states with various transverse momentum (and corresponding energy $E_y$) can be analytically integrated, as it is implemented in nanoMOS, see [38] for details. Therefore, unless otherwise specified, energy *E* in the paper refers to the longitudinal energy due to motion along the *x* direction. In the example mostly used in this paper, the channel length is set at *12nm*, the channel width is *3nm*, and the thickness of both top and bottom gate oxides is *1nm*.

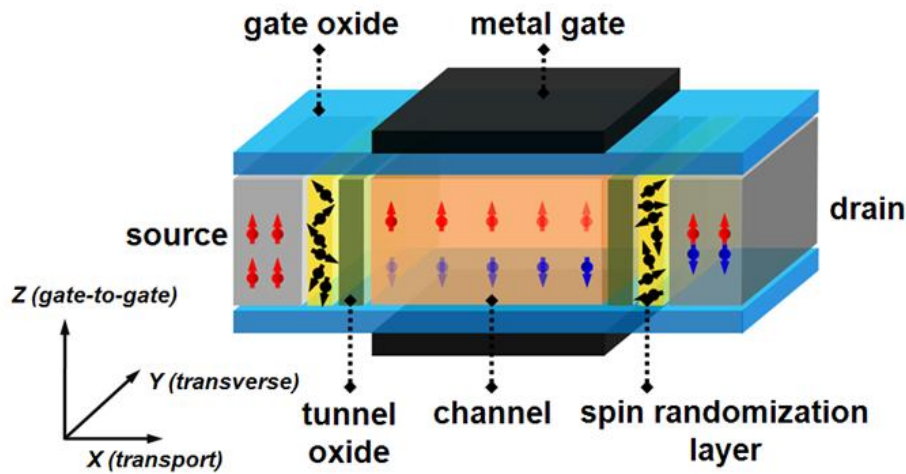

Fig. 1. The spinFET schematic. The source and drain are HMF. The magnetization of the drain can be switched to obtain the parallel and anti-parallel configurations. The double metal gates control the channel electrostatics. The source contact injects and the drain contact detects spin polarized current through an oxide tunneling barrier. A spin randomization layer exists at the boundary of the HMF.

As the NEGF formalism is applied to spinFET, each element in the Hamiltonian $[H]$ is a $2\times2$ matrix, with the $(1,1)$ element representing the onsite energy of "up"-spin state and the $(2,2)$ element – of the "down"-spin state, relative to a chosen preferred axis. Therefore the resulting size of the Hamiltonian matrix $[H]$ is $2N\times2N$. The same is true for the contact self-energy $[\Sigma_{L/R}]$, whose elements are all zeros except for the top-left and the low-right $2\times2$ blocks. The non-zero elements in the contact self-energy describe the coupling of up-spin and down-spin electron states in the source/drain and the channel:

$$\begin{bmatrix} -te^{ik_{L/R,u}a} & 0 \\ 0 & -te^{ik_{L/R,d}a} \end{bmatrix} \qquad (14)$$

where $k_{L/R,u/d}$ is the momentum of the carrier in the source/drain in the up or down spin state, corresponding to the energy $E$ and $t$ is the amplitude of coupling between the source/drain and the channel. We assume that the magnetization of the contacts is along the same preferred axis, otherwise a transformation matrix has to be introduced in the above equation [31]. In the following analysis, the up-spin is set as the majority spin states and down-spin is the minority spin states in the source contact. If the magnetization of the contacts is parallel, the drain contact shares the same spin relation, whereas for the anti-parallel case the drain contact has exactly the opposite relation between up-down and majority-minority designation.

There are two important parameters of the ferromagnetic contacts: the spin splitting $\Delta_S$ and the majority spin bandwidth $E_\omega$. The energy gap between the bottoms of the majority spin band and the bottom of the minority spin band is called the spin splitting $\Delta_S$. The majority spin bandwidth $E_\omega$ is defined as the energy distance between the contact Fermi level $E_{F,L/R}$ and the bottom of the majority spin band. If $\Delta_S$ is larger than $E_\omega$, the ferromagnet is called a half-metal ferromagnet (HMF). The Fermi level crosses just one spin band in such a material. Spin polarization close to 100% is expected in the material if the magnetization is coincident with the up-spin axis, however it does not necessarily translate in extremely high spin polarization in the semiconductor.

We assume that the source and the drain are made of a half-metallic ferromagnet (HMF). The effects of non-ideal spin polarization of carriers are accounted for by the spin randomization layer [17,18], as shown in Fig. 1. It is a layer at the interface of a ferromagnet, where the spins of localized electrons are not aligned with the direction of magnetization, but instead have random directions. The effect of the spin randomization layer is described as the first and the last block in the scattering self-energy $[\Sigma_S]$. The rest of the diagonal blocks in $[\Sigma_S]$ represent spin relaxation in the channel, with, in general, different rates of relaxation.

A tunneling oxide layer may be formed between the source/drain and the channel. It is modeled as a potential barrier of width $W$ and energy height $U_H$. They are the input values of the channel Hamiltonian $[H]$ and set constant at every self-consistent calculation. Since the tunneling barrier has the resistance which is spin-dependent, it is commonly used in ferromagnetic tunnel junctions to increase their magnetoresistance [4]. Drift diffusion simulations [21] predict that the tunneling barrier with carefully adjusted resistance can increase the magnetoresistance of a FM/semiconductor/FM stack as well. This effect exists for any tunneling barrier due to the fact that different states within a band align close to the top of the barrier for up-spin and down-spin bands. However, it is especially pronounced for certain tunneling barrier materials such as MgO [39]. In that case, up-spin and down-spin states over a certain range of energy belong to different bands with different crystal symmetries. Therefore they tunnel with drastically different probabilities. As a result, MgO provides additional very efficient spin filtering and increases the spin polarization of injected carriers. This effect can in principle be modeled by setting different height of the barrier or by different mass of carriers in the barrier for up-spin and down-spin carriers. However, variation of the mass along the transport direction is incompatible with the summation over the transverse momentum states implemented in nanoMOS. Separate NEGF solution for each momentum state is possible [33] but proves to be too computationally expensive in the case of a transistor. In this paper we assume a constant effective mass in the transport direction which is incorporated in nanoMOS.

The explicit form of the Green's functions and self energies with spin indices can be written as a set of diagonal blocks for each grid point

$$G = \begin{pmatrix} G_{uu} & G_{ud} \\ G_{du} & G_{dd} \end{pmatrix}, \quad \Sigma = \begin{pmatrix} \Sigma_{uu} & \Sigma_{ud} \\ \Sigma_{du} & \Sigma_{dd} \end{pmatrix} \quad (15)$$

The in-/out-scattering functions implement the spin-flip scattering processes via the following relation to the electron/hole density $[G^{n/p}]$ and a scattering tensor $[\mathbf{D}]$, see [40]

$$\Sigma_{S,ij}^{in/out}(E) = \sum_{kl} \mathbf{D}_{ijkl}^{n/p}(E) G_{kl}^{n/p}(E) \quad (16)$$

with $[\mathbf{D}]$ being the fourth-order tensors in spin indices at in each grid point. The above equation can be qualitatively understood as the rate of electrons scattering into ($[\Sigma_S^{in}]$) or out of ($[\Sigma_S^{out}]$) the state with energy $E$ being proportional to the existing electron($[G^n]$) or hole($[G^p]$) density.

We assume here the same functional form for electrons $[\mathbf{D}^n]$ and holes $[\mathbf{D}^p]$. The scattering tensor can be separated into the coupling factor and the dimensionless tensor

$$\mathbf{D}(E) = D(E)\Phi \tag{17}$$

For the case of isotropic relaxation, the dimensionless tensor is [40]

$$4\Phi_{ij11} = \begin{pmatrix} 1 & 0 \\ 0 & 2 \end{pmatrix}, \quad 4\Phi_{ij12} = \begin{pmatrix} 0 & -1 \\ 0 & 0 \end{pmatrix}, \quad 4\Phi_{ij21} = \begin{pmatrix} 0 & 0 \\ -1 & 0 \end{pmatrix}, \quad 4\Phi_{ij22} = \begin{pmatrix} 2 & 0 \\ 0 & 1 \end{pmatrix} \tag{18}$$

and the equation for the self-energy turns to,

$$\Sigma_s = \frac{D}{4} \begin{pmatrix} G_{uu} + 2G_{dd} & -G_{ud} \\ -G_{du} & 2G_{uu} + G_{dd} \end{pmatrix} \tag{19}$$

To understand the scattering coupling factor $D$, we now relate it to the commonly used spin life time $T_1$ (or the scattering rate $T_1^{-1}$), which is more familiar to experimentalists, see [4]. For two-dimensional gas of carriers, where the density of states and the spin lifetime are constants of energy, we obtain [38]

$$D(E) = \frac{\hbar^3}{2T_1 m a_x a_y} \tag{20}$$

where $a_x$, $a_y$ are the grid size in $x$ and $y$ directions, and $m$ is the mass of carriers. The spin relaxation rate can be related [21] to the spin diffusion length in a non-degenerate semiconductor with carrier density $n$ and resistivity $\rho$

$$L_s = \sqrt{\frac{T_1 k_B T}{e^2 n \rho}} \tag{21}$$

However for a short channel device, the current is dominated by quantum resistance rather than resistivity of the channel. We will consider cases with widely varying rates of SF scattering. Expected SF times for electrons are of the order of ~0.1ns in silicon and ~1ps in germanium. For holes, SF times are comparable to momentum relaxation times and can be as fast as ~1fs.

## 3. Results

### 3.1 Coherent transport

Let us first consider the case of no spin scattering. In the spinFET studied in this paper, the source Fermi level lies between the majority(up-spin) and minority(down-spin) spin bands, with the parameters spin splitting $\Delta_S$=2.4 and majority spin bandwidth $E_\omega$=2.0 which agrees with theoretical calculation in [18]. The energy difference of 0.4eV between the Fermi level and the minority spin band is big enough to ensure that almost 100% of the injected electrons are up-spin. Absence of scattering will result in a ballistic electron transport [19], i.e. the current reaching the drain end is also 100% up-spin polarized without losing the phase coherence. The carriers see different potential barriers for up-spin and down-spin in the drain contact of different magnetization configurations, and therefore produce totally different I-V characteristics as shown in Fig. 2.

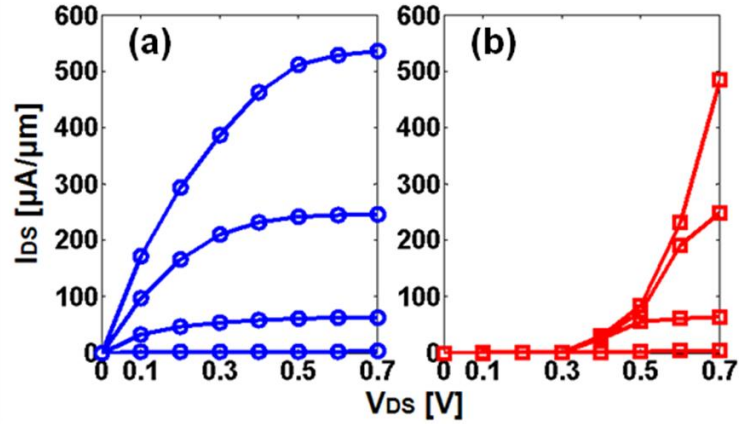

Fig. 2 $I_{DS}$-$V_{DS}$ plots for (a) parallel and (b) anti-parallel configurations in ballistic transport regime. The gate voltage values are *0.7V, 0.5V, 0.3V, 0.1V*, from top to bottom curves.

The up-spin channel is the only conducting channel in the ballistic transport, and it has a high barrier potential in the drain contact under the anti-parallel configuration which blocks the current flow and results in a very small drain current, as can be seen in Fig. 2(b) when $V_{DS}<0.4V$. The *0.4V* is called turn-on voltage $V_{ON}$ here, which is defined as the drain voltage required to push the minority spin band in the drain contact below the Fermi level of the source contact in the anti-parallel configuration. When $V_{DS}>V_{ON}$, the minority spin band has states the Fermi levels of source and drain; and the current will flow, as shown in Fig. 2(b) for $V_{DS}>0.4V$.

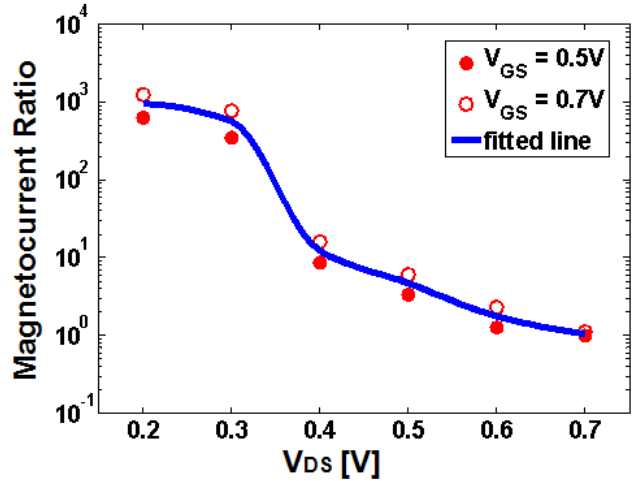

Fig. 3 Magnetocurrent ratio of the ballistic spinFET under different drain bias. The dots are the data obtained as $V_{GS}$=0.5V or 0.7V, and a fitted curve is plotted to represent the average values of the discrete dots.

The MR ratio plotted in Fig. 3 shows that with an ideal ballistic carrier transport a high value of MR around 1000 can be obtained. The lower bound of $V_{DS}$ is chosen in Fig. 3 to ensure the large MR ratio as well as a reasonable drive current of the spinFETs.

### 3.2 Scattering transport

Now let us turn to the effect of SF scattering on the performance of the devices. First, we introduce only the channel SF scattering and leave out the randomization layer and the tunneling barrier, as designated in Fig. 1. The injected 100% up-spin electrons scatter with the SF impurity with some probability and flip into down-spin along the channel. This scattering occurs everywhere inside the channel, as shown in Fig. 4. The closer the electron is to the drain, the higher probability it has to turn into down-spin. The amount of down-spin current increases with the increase of electron-impurity coupling strength $D(E)$. As can be seen in the following analysis, this large amount of down-spin electrons produced in the channel will cause current leakage into the drain and will degrade the device performance.

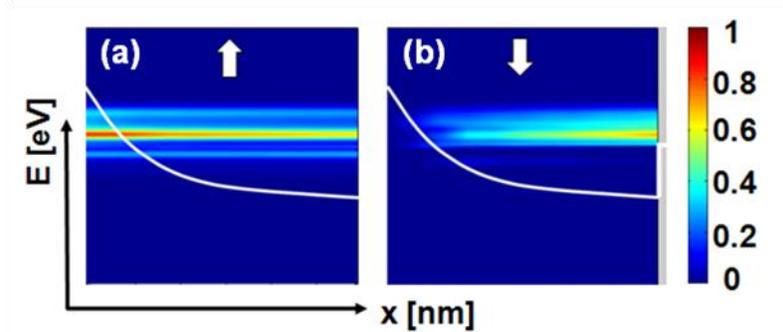

Fig. 4 Energy-position resolved current in the channel for (a) up-spin and (b) down-spin in the parallel configuration. Up-spin turns to down-spin as the electrons traverse the device.

The impurity scattering acts as a cause of shortened lifetime of carriers in the channel. In other words, the local density of state will spread out in real space and broaden in energy space as we increase the value *D(E)*. This effect is observed in our simulation (Fig. 5) for different *D(E)* which corresponding to different spin life times.

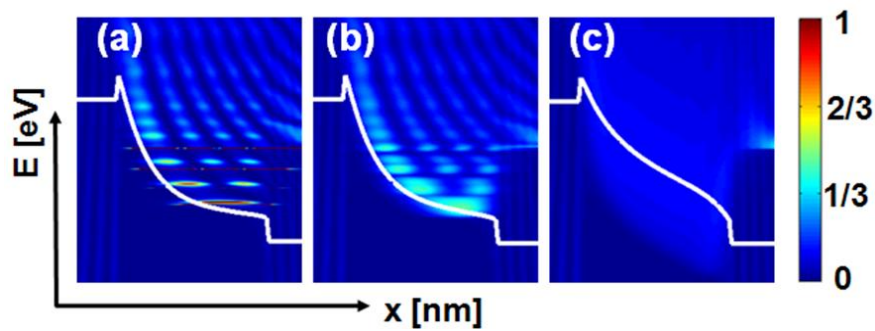

Fig. 5 Energy-position resolved local density of state in the channel in parallel configuration. The electron-impurity coupling *D(E)* are *$2.5 \times 10^{-5}$ $eV^2$* in (a), *$2.5 \times 10^{-3}$ $eV^2$*

in (b) and $1\ eV^2$ in (c), which corresponding to *40ps*, *0.4ps*, and *1fs* spin life times respectively. The strong coupling reduces the spin life time, and also broadens the available states in the channel.

It is also seen in Fig. 5 that the band edge profiles inside the channel are different for these three values of *D(E)*. It can be understood considering that electron distribution for both up-spin and down-spin depends heavily on the SF scattering rate, and thus the modified charge density generates various potential energy profiles according to the Poisson equation. This dependence shows us the importance of a self-consistent solution of NEGF and Poisson equation in the presence of the scattering in the channel. It is inaccurate to assume that the band profiles are the same with and without SF scattering. The charge distribution will affect the energy band, and vice versa. From the above analysis, we conclude that the SF scattering can (1) flip the spin polarizations and create down-spin current along the channel, (2) broaden the local density of states and (3) change the energy profile in the devices.

Besides of the above effects, SF scattering affects MR of the spinFETs. Fig. 6 shows that with SF scattering in the channel, the drain current in the anti-parallel configuration increases dramatically even below the turn-on voltage (compare it to Fig. 2(b)).

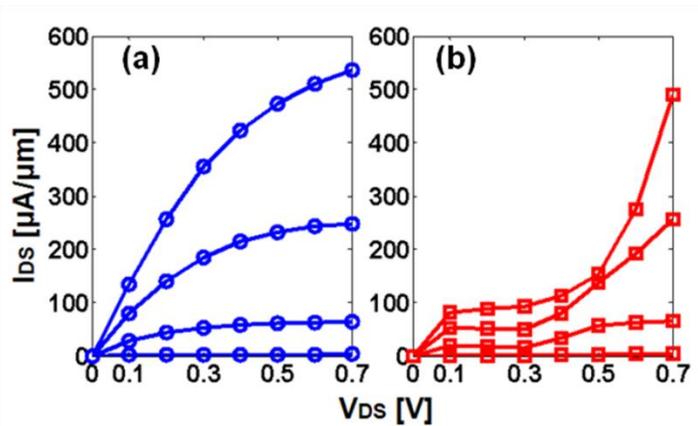

Fig. 6 $I_{DS}$-$V_{DS}$ plots for (a) parallel and (b) anti-parallel configurations in scattering transport regime. The gate voltage values are *0.7V, 0.5V, 0.3V, 0.1V*, from top to bottom curves. The electron-impurity coupling is set to be $10^{-3}eV^2$, which corresponds to *1ps* spin lifetime in the channel.

SF scattering induced leakage in the drain current can greatly decrease the MR of the devices, as shown in Fig. 7 for three different electron-impurity coupling strength corresponding to spin life times of *1ps, 5ps* and *10ps*.

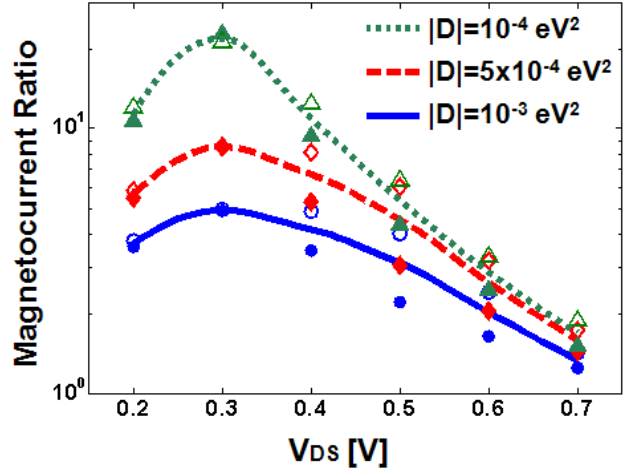

Fig. 7 Magnetocurrent ratio of spinFETs with SF scattering under different drain bias, and with different electron-impurity coupling. The dots are the data obtained as $V_{GS}$=*0.5V* or *0.7V*, and fitted curves are plotted to represent the average values of discrete dots.

Since this drain leakage current in the anti-parallel state is not observed in the ballistic transport (Fig. 2(b)), one must conclude that SF scattering is the cause. Fig. 8 separates up-spin and down-spin currents in the anti-parallel configuration for both ballistic and scattering cases with the bias condition $V_{GS}$=$V_{DD}$=*0.7V*, $V_{DS}$=*0.2V*. Before the device is turned on ($V_{DS}$<$V_{ON}$=*0.4V*) and without SF scattering, almost 100% up-spin electrons injected from the source are confined in the quantum well formed by the channel and cannot escape into the drain (Fig. 8(c)). The negligible down-spin current flows freely from source to the drain, but contributes very little to overall current (Fig. 8(d)). When SF scattering is turned on, a large amount of down-spin electrons is generated (Fig. 8(b)). They escape to the drain contact thanks to the low barrier between the channel and the drain. The up-spin electrons remain confined in the channel as in Fig. 8(a). Note that in the up-spin quantum well, the electrons occupy certain eigenstates of energy. One can notice the 5 lowest modes that contain from one to five anti-nodes of the wavefunction, respectively (Fig. 8(a),(c)). The energy states are wider in the case shown in (a) than in (c), because of the above mentioned SF coupling values.

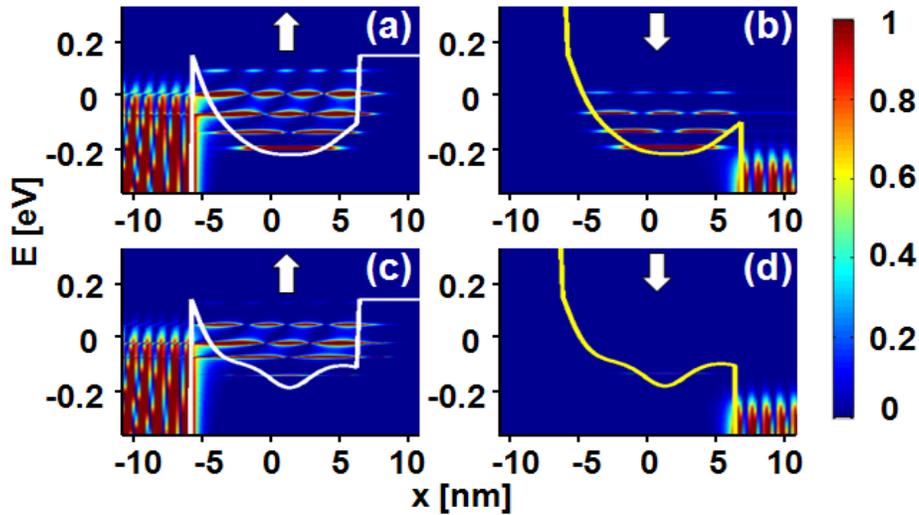

Fig. 8 Energy-position resolved charge density of up-spin ((a) and (c)) and down-spin ((b) and (d)) current, for scattering ((a) and (b)) and ballistic ((c) and (d)) transport regimes. In the scattering transport regimes, the up-spins turn to down-spins and escape to the drain, while no down-spins current flows in the ballistic case.

The interface spin randomization layer can also have the same effect as the channel scattering and be detrimental to the MR ratio. It has been found that the interface treatment at the drain side is more pertinent to achieving high MR ratio [19]. With a reasonable estimate for the coupling strength $D(E)=1eV^2$ and ~80% spin injection polarization, the MR ratio drops drastically compared to the ideal case without the spin randomization layers, as indicated in Fig. 9.

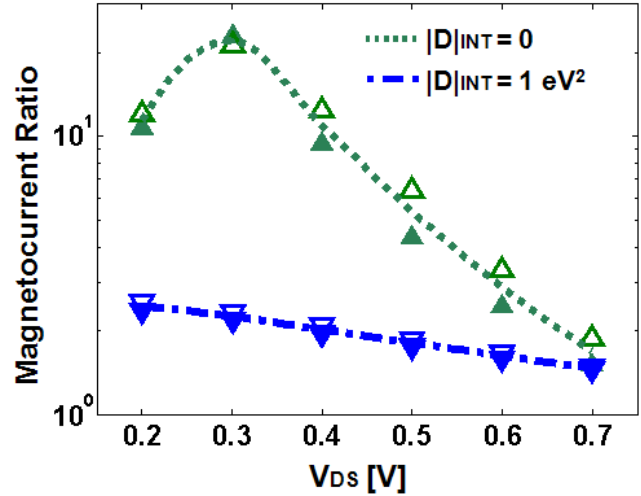

Fig. 9 Magnetocurrent ratio of the spinFETs with SF scattering and interface spin scattering under different drain bias, and with different interface electron-impurity coupling. The channel electron-impurity coupling is $10^{-4}eV^2$ (10ps spin life time). The dots are the data obtained as $V_{GS}=0.5V$ or 0.7V, and fitted curves are plotted to represent the middle point of the discrete dots.

In order to improve the MR ratio, the current in the parallel configuration should be maximized and the current in the anti-parallel configuration should be minimized. In the parallel configuration the down-spin channel is not conductive with or without SF scattering, because the band edge profile contains a high potential wall in the drain end. The up-spin electron transport is similar to that in a Schottky barrier field effect transistor. The comparison of Fig. 2(b) and Fig. 6(b) stresses the need to decrease the current in the anti-parallel configuration as the only way to improve the MR ratio. The magnitudes of up-spin and down-spin current in parallel and anti-parallel configurations are plotted in Fig. 10.

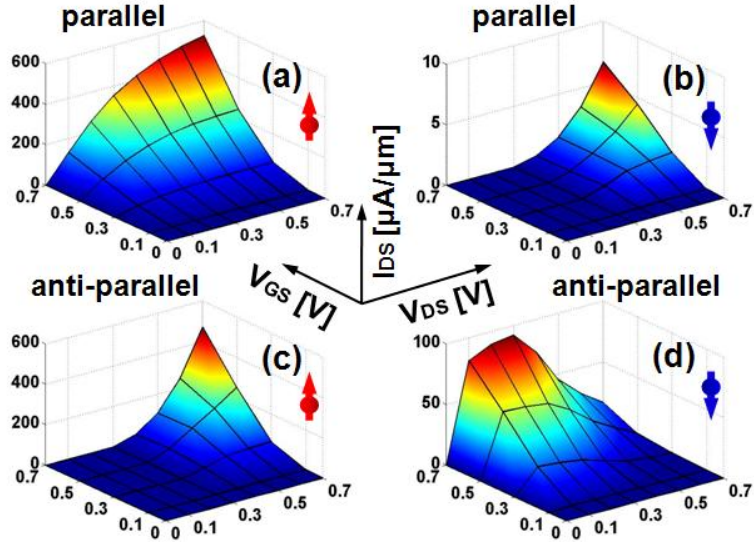

Fig. 10 $I_{DS}$-$V_{DS}$,$V_{GS}$ plots of up-spin ((a) and (c)) and down-spin ((b) and (d)) current, for parallel ((a) and (b)) and anti-parallel configuration ((c) and (d)). The electron-impurity coupling ($|D|=10^{-3}eV^2$) gives large up-spin current at the on state and large down-spin current at medium $V_{DS}$ in the anti-parallel configuration.

The subplots (a) and (b) verify the dominance of the up-spin current in the parallel configuration even at relatively high SF coupling (spin life time of ~1ps). The anti-parallel up-spin current increases to almost *600 µA/µm* in the on-state as seen in Fig. 10(c), which can be explained with the help of the energy-position resolved charge density plot in Fig. 11. In the on-state with $V_{GS}=V_{DS}=V_{DD}=0.7V$, the high gate bias creates thinnest thin Schottky barrier between the source and the channel, permitting a large amount of electrons to tunnel through. The high drain bias ensures that the bottom of the minority carrier conduction band in the drain is below the source Fermi level; and therefore large current flows. However, below the turn-on voltage ($V_{DS}<V_{ON}$), down-spin current due to SF scattering is much larger (Fig. 10(d)) than the up-spin current limited by the quantum well confinement, as seen in Fig.8(b). Thus the up-spin current dominates the total current in the on-state of the spinFETs, and the down-spin current dominate in the off-state.

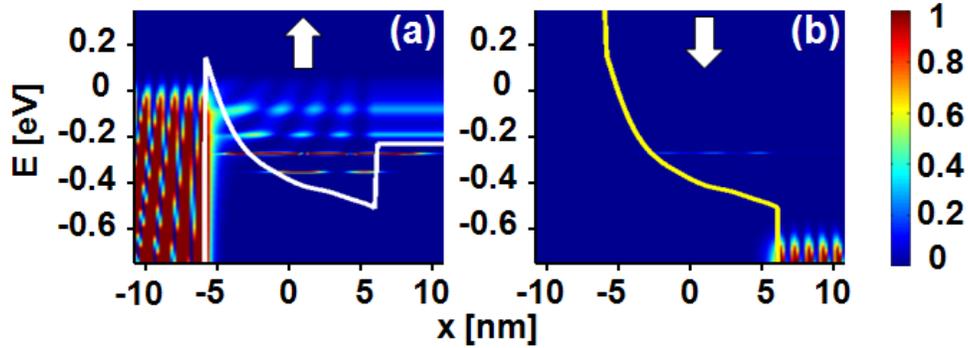

Fig. 11 Energy-position resolved charge density of the (a) up-spin and the (b) down-spin current in the scattering transport at the on-state in the anti-parallel configuration. The high $V_{DS}$ pushes down the drain energy band, which gives a large amount of up-spin current flowing out of drain.

To decrease the high anti-parallel current, two solutions are considered here. The first one is to reduce the up-spin current at $V_{DS}>V_{ON}$ by increasing the spin splitting $\Delta_S$ in the drain contact. The band diagram and charge density are plotted in Fig. 12. The large $\Delta_S$ presents a high potential wall to electrons arriving at the drain and thereby blocks the current. The simulation indicates that the up-spin current is reduced from *560 µA/µm* to *0.6 µA/µm* at the same bias conditions. This is due to the fact that the quantum well confines the up-spin electrons in the channel, increasing the probability of SF scattering into down-spin states. More down-spin electrons are generated in the case of a larger $\Delta_S$ in the drain, and therefore the down-spin current increases from *7 µA/µm* to *57 µA/µm* (Fig. 11(b) and 12(b)). However the total current drops as a result of the dramatic decrease of the up-spin current.

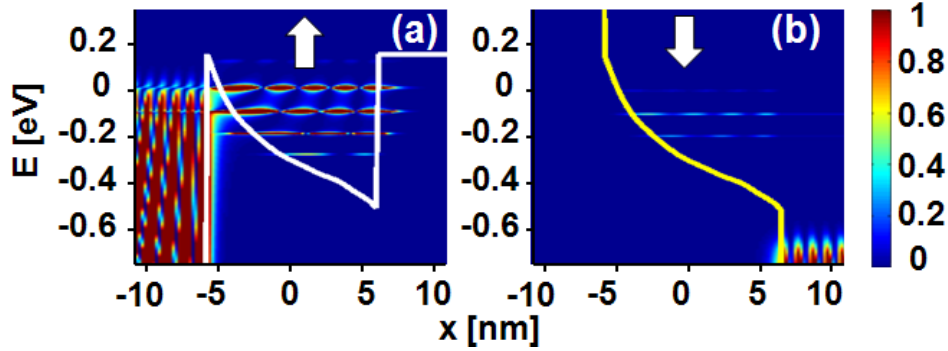

Fig. 12 Energy-position resolved charge density of the (a) up-spin and the (b) down-spin current in the scattering regime at the on-state. The source and drain are anti-parallel configured. The high $V_{DS}$ pushes down the drain energy band, but the large spin splitting blocks the electrons from going into the drain, which reduces the current at the on-state.

The second method is to reduce the down-spin current at $V_{DS}<V_{ON}$ by inserting a tunneling barrier between channel and drain. The high anti-parallel leakage down-spin current at $V_{DS}<V_{ON}$ induced by SF scattering is the main cause of low MR ratio (Fig. 13(b)). The tunneling potential barrier effectively blocks the current and diminishes the leakage, as shown in Fig. 13(a). We simulate a *4nm* thick spin-dependent tunneling barrier that exhibits a higher barrier height for down-spin and a low barrier height for up-spin electrons. In the parallel configuration, the up-spin dominated current changes insignificantly, while the down-spin leakage current in anti-parallel configuration is lower. The effect of the spin-dependent tunneling oxide is exhibited at both source and drain ends. The barrier at the source end can filter the injected current and increase its polarization, and the barrier at the drain end can stop the leakage down-spin current below the turn-on voltage and almost eliminate the current in the anti-parallel configuration. Thus the MR ratio is ~*100x* higher with the spin selective tunneling oxide than without it.

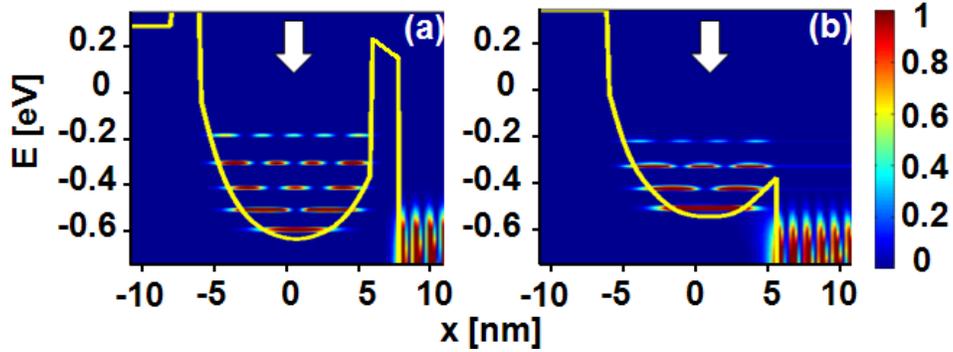

Fig. 13 Energy-position resolved charge density of the down-spin current in the scattering regime. The source and drain are anti-parallel configured. The tunneling barrier for the down-spin electrons between the channel and drain can lower the total current and therefore increase MR by about *100x*.

The enhancement of MR ratio by adding the tunneling barriers can be seen in Fig. 14. In the on-state that $V_{GS}=V_{DD}=0.7V$, the spinFETs without the tunneling barriers have a low MR of ~*20*. It can reach *100* with the insertion of the same barriers ($U_{HD}=U_{HU}$) for both up- and down-spins. In the case that the tunneling barrier for the down-spin is higher than that for the up-spin electrons ($U_{HD}>U_{HU}$), the MR can increase to ~*500*.

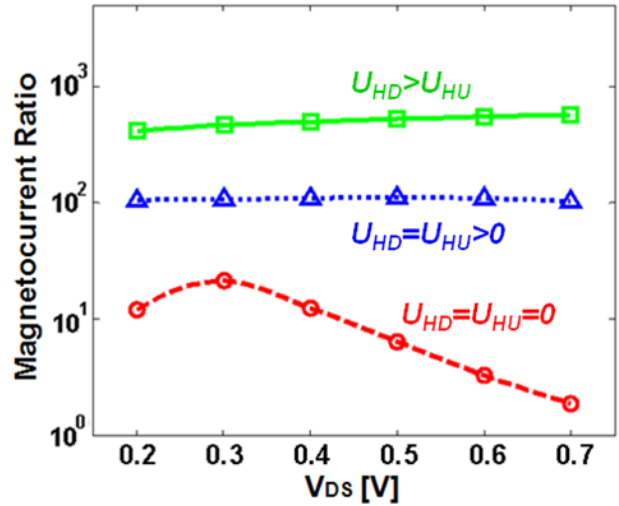

Fig. 14 Magnetocurrent ratio of the spinFETs with various tunneling barriers configurations under different drain bias. The tunneling barriers in the source and drain ends are of *4nm* thick and *0.6eV* high. There are 3 devices simulated here: without tunneling barriers for both up- and down-spins (dash line), with the same tunneling barriers for both spins (dotted line), and with the different barriers for both spins (solid line).

## 4. Conclusion

In this work we have demonstrated a rigorous quantum transport (NEGF) simulation of spinFET with the account of SF scattering, tunneling and Schottky barriers, and self-consistent charge distribution. In the ideal case without channel scattering the device shows very large MR ratio of the order of $10^3$. SF scattering generates a large amount of down-spin electrons which increases the current in the anti-parallel configuration, and eventually degrades the MR ratio to around 10 with a reasonable assumption of spin life time in Silicon. The MR ratio becomes even lower with the account of the inevitable spin randomization layer at the FM/semiconductor interface. As a result of our numerical study, two solutions are proposed to improve the performance of spinFETs. The first method is to increase the energy spin splitting in the drain contact in order to create a high potential barrier to block the drain leakage current, which mainly consist of the up-spin electrons coming from the source. Another solution is to insert a spin-selective tunneling oxide layer between the source/drain and the channel, which brings the MR ratio up to ~500.

## 5. Acknowledgements

The authors gratefully acknowledge the support of the Nanoelectronic Research Initiative and the Network for Computational Nanotechnology. We express our appreciation of helpful discussions with Supriyo Datta, Brian Doyle, and George Bourianoff.

# 6. References

[1] Semiconductor Industry Association, International Technology Roadmap for Semiconductors, Chapter "Emerging Research Devices", http://public.itrs.net/ (2009).

[2] S. A. Wolf, D. D. Awschalom, R. A. Buhrman, J. M. Daughton, S. von Molnar, M. L. Roukes, A. Y. Chtchelkanova, and D. M. Treger, Science 294, 1488 (2001).

[3] A. Fert, Rev. of Mod. Phys. 80, 1517 (2008).

[4] I. Zutić, J. Fabian, and S. D. Sarma, Rev. of Mod. Phys. 76, 323 (2004).

[5] M. N. Baibich, J. M. Broto, A. Fert, F. Nguyen Van Dau, F. Petroff, P. Etienne, G. Creuzet, A. Friederich, and J. Chazelas, Phys. Rev. Lett. 61, 2472 (1988).

[6] G. Binash, P. Grünberg, F. Saurenbach, and W. Zinn, Phys. Rev. B 39, 4828 (1989).

[7] S. S. P. Parkin, C. Kaiser, A. Panchula, P. M. Rice, B. Hughes, M. Sament, and S. Yang, Nature Mater. 3, 862 (2004).

[8] Myers, E. B., Ralph, D. C., Katine, J. A., Louie, R. N. & Buhrman, R. A., Science 285, 867–870 (1999).

[9] S. Sugahara and M. Tanaka, Appl. Phys. Lett. 84, 2307 (2004).

[10] B. Huang, D. J. Monsma, and I. Appelbaum, Phys. Rev. Lett. 99, 177209 (2007).

[11] S. S. P. Parkin, K. P. Roche, M. G. Samant, P. M. Rice, R. B. Beyers, and R. E. Scheuerlein, J. of Appl. Phys. 85, 5828 (1999).

[12] S. Datta and B. Das, Appl. Phys. Lett. 56, 665 (1990).

[13] R. A. de Groot, F. M. Mueler, P. G. van Eugen, and K. H. J. Buschow, Phys. Rev. Lett. 50, 2024 (1983).

[14]  R. B. Mancoff, B. M. Clemens, E. J. Singley, and D. N. Basov, Phys. Rev. B 60, R12565 (1999).

[15]  K. Inomata, N. Ikeda, N. Tezuka, R. Goto, S. Sugimoto, M. Wojcik, and E. Jedryka, Sci. Tech. Adv. Mater. 9, 014101 (2008).

[16]  M. Tanaka and S. Sugahara, IEEE Trans. on Elec. Dev. 54, 961 (2007).

[17]  C. T. Tanaka, J. Nowak, and J. S. Moodera, J. of Appl. Phys. 86, 6239 (1999).

[18]  G. A. de Wijs and R. A. de Groot, Phys. Rev. B 64, 020402 (2001).

[19]  T. Low, M. S. Lundstrom, and D. E. Nikonov, J. of Appl. Phys. 104, 094511 (2008).

[20]  G. Schmidt, D. Ferrand, and L. W. Molenkamp, Phys. Rev. Lett. 62, R4790 (2000).

[21]  Rashba, E. I., Phys. Rev. B 62, R16 267–R16 270 (2000).

[22]  A. Fert and H. Jaffrès, Phys. Rev. B 64, 184420 (2001).

[23]  H. C. Koo, H. Yi, J.-B. Ko, J. Chang, S.-H. Han, D. Jung, S.-G. Huh, and J. Eom, Appl. Phys. Lett. 90, 022101 (2007).

[24]  X. Lou, C, Adelmann, S. A. Crooker, E. S. Garlidi, J. Zhang, K. S. Madhukar Reddy, S. D. Flexner, C. J. Palmstrøm, and P. A. Crowell, Nature Physics 3, 197 (2007).

[25]  W. Van Roy, P. Van Dorpe, R. Vanheertum, P.-J. Vandormael, and G. Borghs, IEEE Trans. Electron. Devices 54, 933 (2007).

[26]  B.-C. Min, K. Motohashi, C. Lodder, and R. Jansen, Nature Materials 5, 817 (2006).

[27]  Y. Zhou, M. Ogawa, M. Bao, W. Han, R. K. Kawakami, and K. Wang, Appl. Phys. Lett. 94, 242104 (2009).

[28]  T. Marukame, T. Inokuchi, M. Ishikawa, H. Sugiyama, and Y. Saito, IEDM Technical Digest, 09-2 (2009).

[29]  A. M. Bratkovsky, Rep. Prog. Phys. **71,** 026502 (2008).

[30]   I. Zutić, J. Fabian, and S. C. Erwin, IBM J. Research & Development 50, 121 (2006).

[31]   S. Datta, Quantum Transport: Atom to Transistor, Cambridge University Press (2005).

[32]   A. A. Yanik, G. Klimeck, and S. Datta, Phys. Rev. B 76, 045213 (2007).

[33]  S. Salahuddin, D. Datta, P. Srivastava, S. Datta, IEEE Int. Elec. Dev. Meeting Technical Digest, p. 121 (2007).

[34] X.-G. Zhang and W. H. Butler, Phys. Rev. B 70, 172407 (2004).

[35]  D. Waldron, L. Liu, and H. Guo, Nanotechnology 18, 424026 (2007).

[36]   Z. Ren, S. Goasguen, A. Matsudaira, S. S. Ahmed, K. Cantley, and M. Lundstrom. (2006). nanoMOS. [Online]. Available: https://www.nanohub.org/tools/nanomos/

[37]   M. P. Anantram, M. S. Lundstrom, and D. E. Nikonov, Proceedings of the IEEE, 96, 1511 - 1550 (2008).

[38]   D. E. Nikonov, G. I Bourianoff, and P. A. Gargini, http://nanohub.org/resources/7772.

[39]   W.H. Butler and A. Gupta, Nature Materials, 4, 845-847 (2004).

[40]   S. Datta, Proc. of Inter. School of Phys. Societa Italiana di Fisica, unpublished, p. 244 (2004).